\documentclass[aps,prl,groupedaddress,twocolumn,showpacs,floatfix]{revtex4}

\usepackage{graphicx}
\usepackage{amsmath}
\usepackage{psfrag}
\usepackage{natbib}

\begin{document}
\preprint{}

\title{Self-Assembly of Monatomic Complex Crystals and Quasicrystals\\
  with a Double-Well Interaction Potential}

\author{Michael Engel}
\email{mengel@itap.uni-stuttgart.de}
\author{Hans-Rainer Trebin}
\affiliation{Institut f\"ur Theoretische und Angewandte Physik, Universit\"at
  Stuttgart, Pfaffenwaldring 57, 70550 Stuttgart, Germany}

\begin{abstract}
  For the study of crystal formation and dynamics we introduce a simple
  two-dimensional monatomic model system with a parametrized interaction
  potential. We find in molecular dynamics simulations that a surprising
  variety of crystals, a decagonal and a dodecagonal quasicrystal are
  self-assembled. In the case of the quasicrystals the particles reorder by
  phason flips at elevated temperatures. During annealing the entropically
  stabilized decagonal quasicrystal undergoes a reversible phase transition at
  $65\%$ of the melting temperature into an approximant, which is monitored by
  the rotation of the de Bruijn surface in hyperspace.
\end{abstract}


\maketitle

Self-assembly is the formation of complex patterns out of simple constituents
without external interference. It is a truly universal phenomenon, fundamental
to all sciences~\cite{Whitesides02}. Although usually the constituents
interact only locally, the result is well-ordered over long distances,
sometimes with a high global symmetry. In the process of crystallization,
particles (atoms, molecules, colloids, etc.) arrange themselves to form
periodic or quasiperiodic structures. Here we are interested in structurally
complex phases. Examples are metallic crystals with large unit cells --
hundreds or thousands of atoms -- known as complex metallic
alloys~\cite{Turchi05}. Some consequences of the complexity are the existence
of an inherent disorder and the formation of well-defined atomic
clusters~\cite{Urban04}. Related alloys differ by the cluster arrangement. In
the limit of infinitely large unit cells non-periodic order like in
quasicrystals~\cite{Janot} is obtained. However self-assembly of complex
phases is not unique to alloys. Recently micellar phases of dendrimers were
observed to form a mesoscopic quasicrystal~\cite{Zeng04}, and there are
indications that quasicrystals exist in monodisperse colloidal (macroscopic)
systems~\cite{Denton98}. Since the interaction between colloidal particles can
be tuned in various ways, these systems are well-suited for experiments
investigating self-assembly in dependence of the potential shape.

All of the previous examples have in common that the crystal growth can be
modeled with effective pair potentials, which is a prerequisite for simulating
self-assembly of a large number of particles on a computer. The first such
simulations were conducted by Lan\c{c}on and Billard in
two~\cite{LanconBillard} and by Dzugutov in three
dimensions~\cite{Dzugutov92}. In the latter work the system was chosen
monatomic to facilitate computation and separate chemical from topological
ordering. It is well-known that many common pair-potentials like the
Lennard-Jones (LJ) potential favor close-packed ground states. To force the
formation of alternative structures the Dzugutov potential is endowed apart
from its LJ minimum with an additional maximum. Although the potential was
originally tailored to lead to a glassy state~\cite{Dzugutov92}, it stabilizes
the $\sigma$-phase at low temperature~\cite{Roth00}. Later, similar potentials
were used to demonstrate the formation of a dodecagonal quasicrystal in two
dimensions~\cite{Quandt99}, which on closer inspection was identified as a
periodic approximant. In fact it is often difficult to distinguish
quasicrystals and closely related periodic complex crystals due to their
structural similarity.

\begin{figure}
  \centering
  \includegraphics[width=0.7\linewidth]{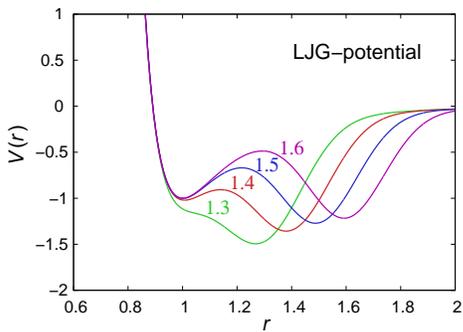}
  \caption{(color online) LJG-potential for $\epsilon=1.1$, $\sigma^{2}=0.02$,
    and $r_{0}=1.3$ (Hex2 in Fig.~\ref{fig:diagram_sim}), $1.4$ (Sqa), $1.5$
    (Pen), $1.6$ (Hex1).\label{fig:ljg_potential}}
\end{figure}

The relation between an interaction potential and its energetic ground state
is a fundamental problem of physics. It can be approached by two methods: The
direct method starts from a given parametrized set of potentials and studies
the resulting structures as a function of the parameters (and
temperature/pressure). An example is the hard core plus linear ramp model with
the ramp slope as single parameter~\cite{Jagla98}. The inverse method tries to
find an appropriate potential that stabilizes a given structure via
optimization~\cite{Rechtsman06}. It was used recently by Rechtsman et al. to
find potentials for various lattices~\cite{RechtsmanDiverse}. In this Letter
we apply the direct method to a potential of the form \cite{Rechtsman06}
\begin{equation}\label{eq:ljg}
V(r)=\frac{1}{r^{12}}-\frac{2}{r^{6}}-
\epsilon\exp\left(-\frac{(r-r_{0})^{2}}{2\sigma^{2}}\right),
\end{equation}
which we denote Lennard-Jones-Gauss (LJG) potential. For most values of the
parameters $V(r)$ is a double-well-potential with the second well at position
$r_{0}$, of depth $\epsilon$ and width $\sigma$
(Fig.~\ref{fig:ljg_potential}). We note that the general form of pair
potentials in metals consists of a strongly repulsive core plus a decaying
oscillatory (Friedel) term~\cite{Friedel}. A LJG-potential can be understood
as such an oscillatory potential, cut off after the second minimum.

In the following we restrict the parameter space by fixing $\sigma^{2}=0.02$.
The $T=0$ phase diagram in the $r_{0}$-$\epsilon$-plane is determined in two
steps: First, candidate ground state structures are obtained from annealing
simulations. Next, a defect free sample of each candidate structure is relaxed
with a conjugate gradient algorithm. During the relaxation particle movements
and adjustments of the simulation box dimensions are allowed. The stable
structures (within the candidates) are the ones with the lowest potential
energies. Simulations were carried out by solving the equations of motion with
molecular dynamics. Periodic boundary conditions and a Nos\'e-Hoover
thermo-/barostat for constant temperature $T$ and constant pressure $P=0$ were
used. A cut-off was applied to the LJG-potential at $r=2.5$.

\begin{figure}
  \centering
  \includegraphics[width=0.95\linewidth]{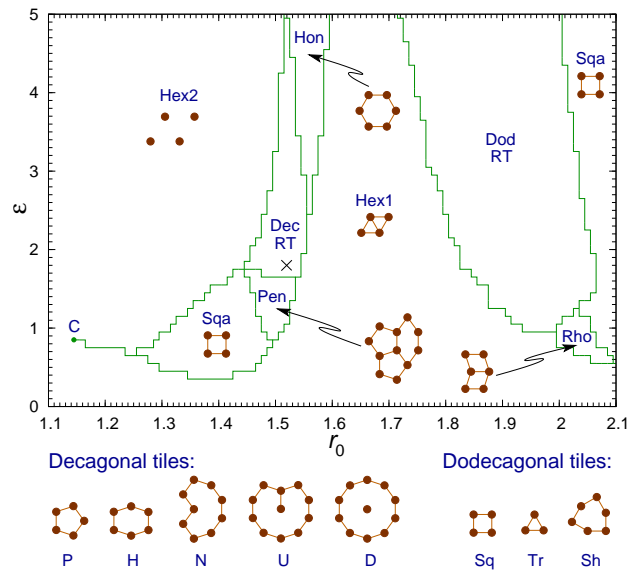}
  \caption{(color online) Phase observation regions after annealing
    simulations. Tilings of the unit cells are shown. The quasicrystal tiles
    are: (P)entagon, (H)exagon, (N)onagon, (U)-tile, (D)ecagon, (Sq)are,
    (Tr)iangle, (Sh)ield.\label{fig:diagram_sim}}
\end{figure}

We performed 5000 annealing simulations with a sample of 1024 particles using
parameters located on a fine grid: $r_{0}\in[1.11,2.10]$, $\Delta r_{0}=0.01$
and $\epsilon\in[0.1,5.0]$, $\Delta\epsilon=0.1$. During each run the
temperature was lowered linearly over $2\cdot 10^{6}$ molecular dynamics steps
starting from the liquid state. A typical final particle configuration
consists of several well-ordered grains whose structure is analyzed in direct
and in reciprocal space. The phase observation regions are displayed in
Fig.~\ref{fig:diagram_sim}. We found hexagonal phases with nearest-neighbor
distance close to 1.0 (Hex1) and close to $r_{0}$ (Hex2), a square lattice
(Sqa), a decagonal (Dec) and a dodecagonal (Dod) quasiperiodic random tiling
(RT), a phase built from pentagons and hexagons (Pen), a rhombic lattice
(Rho), and finally a honeycomb lattice (Hon). It can be seen in
Fig.~\ref{fig:diagram_sim} that the two hexagonal phases are connected around
the phase Sqa.  Across the line between C and the phase Sqa there is a rapid
increase in the hexagonal lattice constant.

\begin{figure}
  \centering
  \includegraphics[width=0.95\linewidth]{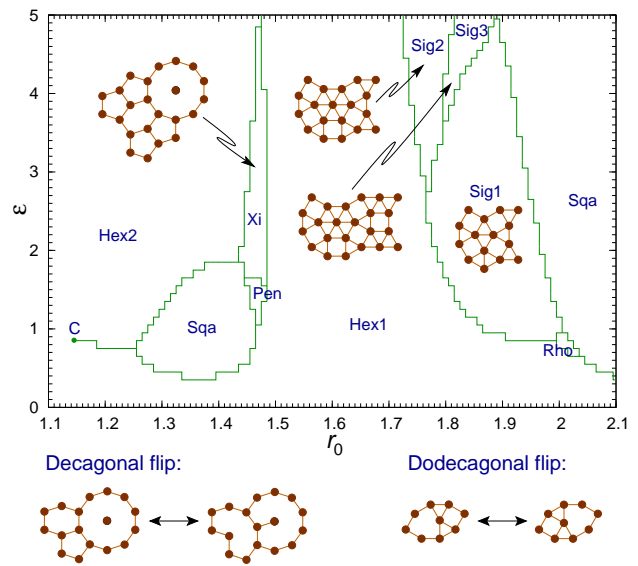}
  \caption{(color online) Phase diagram of the LJG-potential at $T=0$ with
    $\sigma^{2}=0.02$.  Four approximants have been found. Phason flips
    correspond to local changes in the tilings.\label{fig:diagram_calc}}
\end{figure}

The phases resulting from the annealing simulations are possible energy ground
states.  Further possibilities are approximants, which we constructed from the
tiles in Fig.~\ref{fig:diagram_sim}. The choice of approximants is restricted
by the following observations at low temperature: (i)~The D-tiles have lowest
energy and their number is maximized (see below). (ii)~The Sh-tiles and two
neighboring Sq-tiles are avoided. Together, the approximants and the phases
from annealing simulations were used as initial structures for numerical
relaxation. In the phase diagram (Fig.~\ref{fig:diagram_calc}) five complex
crystals are stable: the phase Pen, the decagonal approximant (Xi), which is a
periodic stacking involving D-tiles, and three dodecagonal approximants: the
$\sigma$-phase (Sig1) and two modifications (Sig2), (Sig3). Here we use the
term ``complex'' since the lattice constants are larger than the potential
cut-off, which means that the unit cells have to be stabilized indirectly by
geometric constraints.  Quasicrystals are not energetically stable at $T=0$.
The phase boundaries are slightly displaced compared to those from annealing
simulations due to metastability.

The location of the stability regions can be understood from the near-neighbor
configuration. Local $n$-fold order is stabilized for $r_{0}\approx
2\cos(\pi/n)$. Fig.~\ref{fig:diagram_calc} confirms this except for local
five-fold order, that is found at $r_{0}\approx 1.47$ as a result of the
competition with the close-packed phase Hex1. Structure details of the complex
phases are collected in Tab.~\ref{tab:comparison}: The phases Xi and Dec have
a surprisingly low density, which is only $65\%$ of the density of Hex1.  They
are locally very similar, their potential energies differ by less than
$0.7\%$.

\begin{table}
  \tabcolsep3.0mm
\begin{center}
\begin{tabular}{cccc}\hline\hline
Phase & Density & Lattice constants & Atoms/u.c.\\\hline
Pen  & 0.8981 & $a=2.62$, $b=2.50$ &  5 \\\hline
Xi   & 0.7617 & $a=4.24$, $b=4.24$ & 13 \\
Dec  & 0.7608 & ---                & ---\\\hline
Sig1 & 1.0718 & $a=2.73$, $b=2.73$ &  8 \\
Sig2 & 1.0788 & $a=3.73$, $b=2.73$ & 11 \\
Sig3 & 1.0829 & $a=4.73$, $b=2.73$ & 14 \\
Dod  & 1.0774 & ---                & ---\\\hline\hline
\end{tabular}
\caption{Structure details of the ideal tilings of the complex crystals. For
  comparison the quasicrystals are included.\label{tab:comparison}}
\end{center}
\end{table}

\begin{figure}
  \centering
  \includegraphics[angle=90,width=0.80\linewidth]{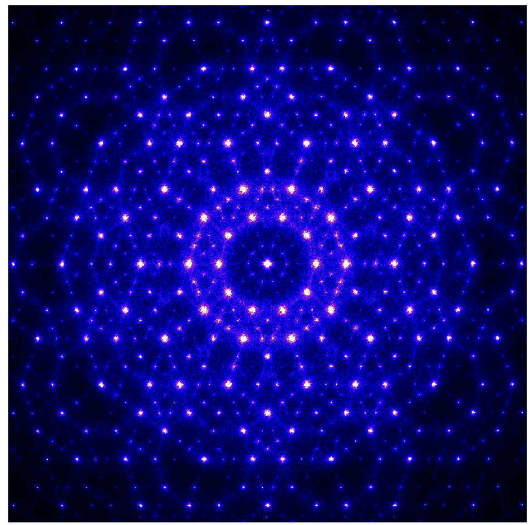}
  \caption{(color online) Diffraction image of the large 10000 particle sample
    with the decagonal quasicrystal. \label{fig:diffraction}}
\end{figure}

We have studied the decagonal quasicrystal in longer simulations using the
parameters $r_{0}=1.52$, $\epsilon=1.8$, and $\sigma^{2}=0.02$ (indicated by a
cross in Fig.~\ref{fig:diagram_sim}).  With these parameters, the decagonal
quasicrystal is assembled with few defects at elevated temperatures. A
simulation of a large sample, 10000 particles, was initiated in a random
configuration at $T=0.50$, which is close to the melting point
$T_{M}=0.56\pm0.02$. In the following, the periodic boundary conditions are
turned off to allow phason strain relaxation. At the beginning the particles
quickly formed an amorphous state with local decagonal order. After about
$10^{5}$ molecular dynamics steps multiple grains with the quasicrystal
started to grow. The bigger ones increased their size until at ca. $10^{7}$
steps only one single grain remained. We continued the simulation up to
$5\cdot 10^{7}$ steps, healing out point defects (vacancies, interstitials)
and improving the quasiperiodicity.  At the end the sample was quenched to
$T=0$ and relaxed.  The diffraction image (Fig.~\ref{fig:diffraction}) shows a
perfect decagonal symmetry with long-range order. There is a weak pattern of
intrinsic diffuse scattering due to the randomness.

In the simulation the dynamics is dominated by particle jumps over the short
distance $\Delta r=0.6$, called phason flips, which transform energetically
comparable configurations into another (see Fig.~\ref{fig:diagram_calc}). Each
such configuration can be mapped to a tiling and embedded as a discrete de
Bruijn surface in a five-dimensional hypercubic lattice by noting that the
tiling vertices are integer multiples of the five basis vectors
$\boldsymbol{e}_{n}=(\cos(2\pi n/5),\sin(2\pi n/5))$. Although the average
orientation of the surface is fixed by the decagonal symmetry, phason flips
lead to local fluctuations $\boldsymbol{h}^{\perp}(\boldsymbol{r})$ in
``perpendicular space'' resulting in a phason strain
$\chi_{ij}=\nabla_{i}h_{j}^{\perp}$. The ensemble of all accessible
configurations forms a entropically stabilized random tiling with a phason
elastic free energy density of the general form
$f(T,\chi)=\lambda_{1}(T)\chi_{1}^{\;2}+\lambda_{2}(T)\chi_{2}^{\;2}$, where
$\chi_{1}^{\;2}$, $\chi_{2}^{\;2}$ are quadratic forms in $\chi_{ij}$, and
$\lambda_{1}$, $\lambda_{2}$ independent phason elastic
constants~\cite{Widom89}.

\begin{figure}
  \centering
  \includegraphics[width=0.80\linewidth]{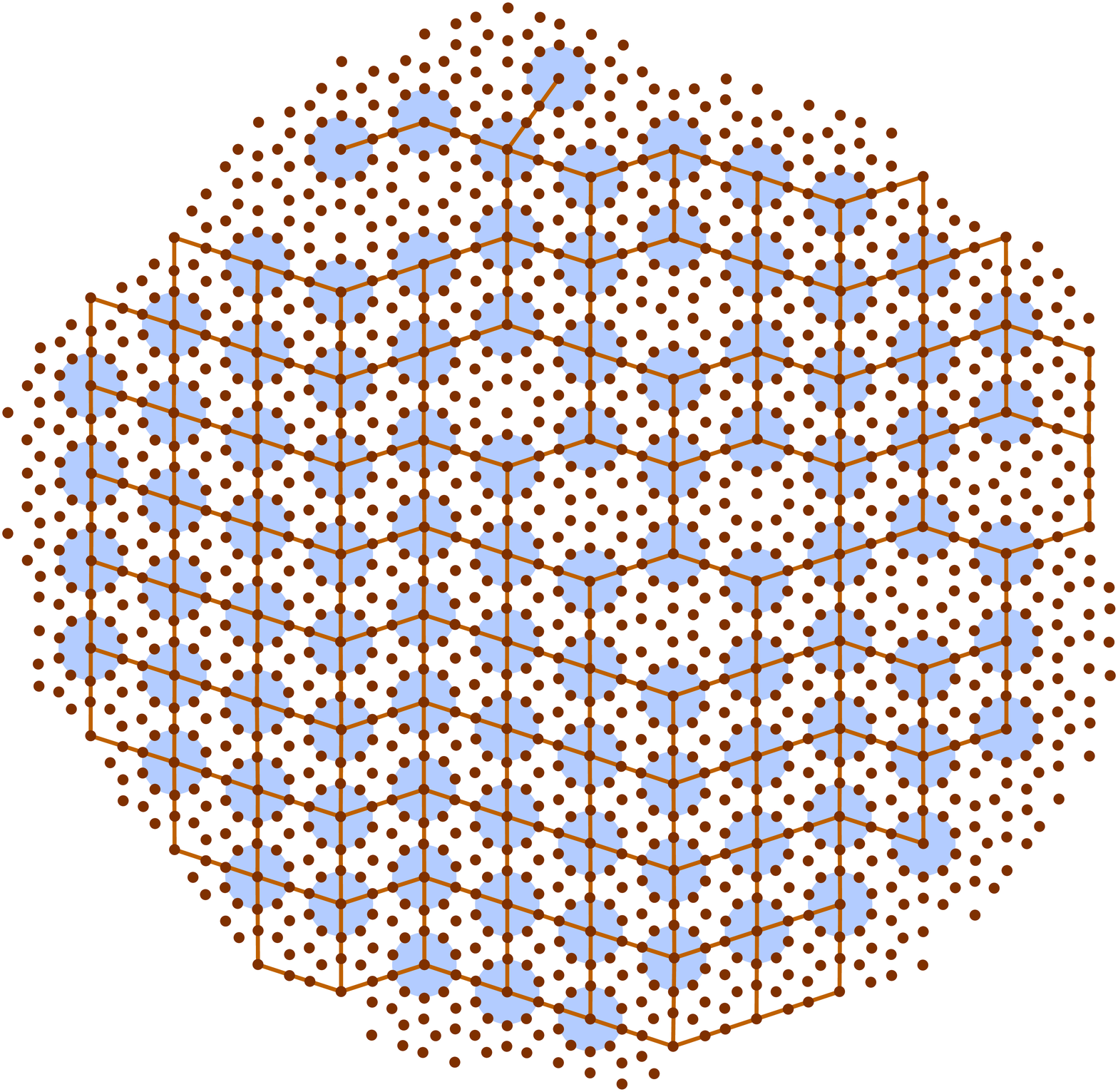}
  \caption{(color online) Particle configuration in molecular dynamics
    simulation at $T=0.30$. The D-tiles are arranged preferably in a rhomb
    super-tiling.\label{fig:tiling}}
\end{figure}

According to the $T=0$ phase diagram a transformation to the phase Xi can
occur during annealing \cite{Lee01}. This is achieved by (i) a collective
rearrangement of the tiling induced by a global change of the de Bruijn
surface orientation and (ii) a damping of the local phason fluctuations. For
(i) to happen a huge number of phason flips is necessary, which makes the
transition extremely slow. We performed long simulations over $10^{9}$ steps
with 1600 particles. At intermediate temperatures, $T<0.40$, a reversible
change in the tiling was found: The density of D-tiles increased and they
arranged preferably close-packed in rhombs, characteristic for the approximant
Xi. However as shown in Fig.~\ref{fig:tiling}, some defects and stacking
faults were still present in the rhomb super-tiling.

\begin{figure}
  \centering
  \includegraphics[width=0.95\linewidth]{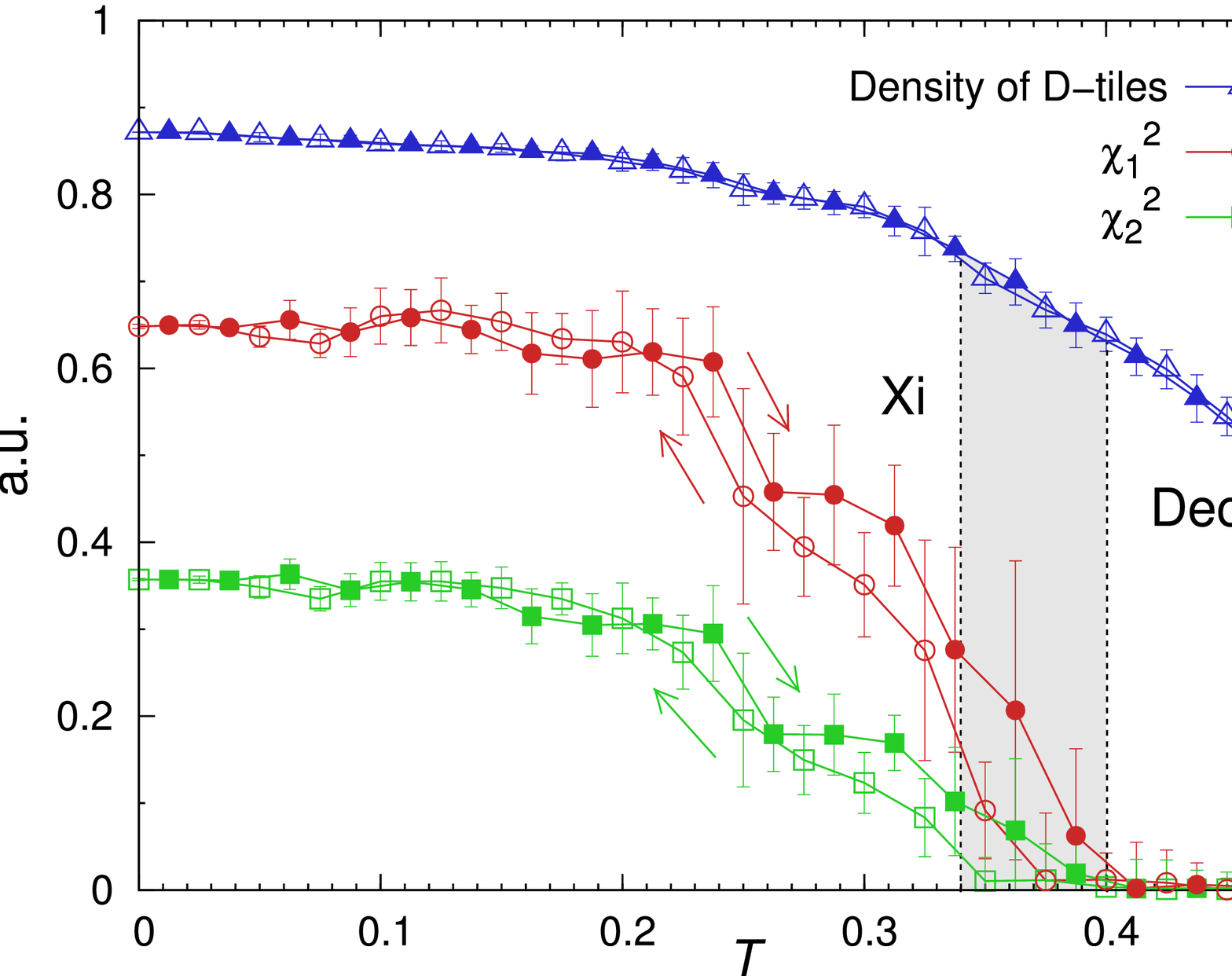}
  \caption{(color online) Monte Carlo simulation of the phase transition
    between the decagonal RT and the approximant Xi. The data has been
    averaged over intermediate temperature intervals. Open symbols: cooling;
    full symbols: heating.\label{fig:transition}}
\end{figure}

At low temperatures, $T<0.30$, the flip frequency was too slow to reach
equilibrium with molecular dynamics. Hence we turned to a Monte Carlo
algorithm, which allowed sampling the full temperature range from $T=0.5$ down
to zero and back up. As elementary step a random displacement inside a circle
of radius 0.7 was used.  The large radius allows both local relaxation and
phason flips.  The squared average phason strains $\chi_{1}^{\;2}$,
$\chi_{2}^{\;2}$ are indicators of the de Bruijn surface orientation and thus
order parameters for the phase transition.  The results in
Fig.~\ref{fig:transition} show a reversible transition at $T_{c}=0.37\pm0.03$.
Above $T_{c}$ there is a strong increase in the density of D-tiles, which then
slows down below $T_{c}$. The phason strains $\chi_{1}^{\;2}$ and
$\chi_{2}^{\;2}$ fluctuate and switch from zero average at $T>T_{c}$ to finite
values including a small hysteresis. We note that with periodic boundary
conditions or in large samples phason flips alone cannot change the decagonal
symmetry efficiently.

In the case of the Sig-phases, phason flips are not possible. They only occur
in combination with Sh-tiles as depicted in Fig.~\ref{fig:diagram_calc}. Even
though Sh-tiles are not seen in the ground states, they are present in
equilibrium at higher temperature as structural defects. The flip mechanism
then differs from the one in the decagonal RT~\cite{Oxborrow93}.

Finally we comment on the value of the parameter $\sigma$ and on other
double-well potentials. The dynamics of the complex phases is controlled by
the potential hill between the minima. A high potential hill leads to a low
phason flip frequency and slow phase transitions, at least in molecular
dynamics. On the other hand, a too low potential hill does not stabilize
complex phases. The phase behavior is quite robust against small changes in
the potential. Phase diagrams resembling Fig.~\ref{fig:diagram_calc} have been
obtained for different values of $\sigma$. Choosing $\sigma^{2}=0.02$
constitutes a compromise between high flip frequency and stability. In
contrast, the flip frequency of earlier models~\cite{LanconBillard,
  Dzugutov92} is much lower. Another example, a repulsive term plus two
negative Gaussians has a qualitatively similar phase diagram, although
additional phases appear. Further details will be presented elsewhere.

In conclusion, we have shown that systems with two competing nearest-neighbor
distances can have a much more complicated phase behavior than what is known
for single minimum potentials. Quasicrystals and complex crystals appear
naturally in such systems as an attempt to maximize local particle
configurations with non-crystallographic symmetry. In the presence of phason
flips, entropic contributions to the free energy play an important role in the
thermodynamic stability.

One of us (M.E.) would like to thank T.\ Odagaki for the hospitality during a
stay at Kyushu University where part of this work was done. Financial support
from the Deutsche Forschungsgemeinschaft under contract number TR 154/24-1 is
gratefully acknowledged.


\begin{thebibliography}{20}

\bibitem{Whitesides02}
G.\ M.\ Whitesides and B.\ Grzybowski, Science {\bf 295}, 2418 (2002).

\bibitem{Turchi05}
P.\ Turchi and T.\ Massalski (editors), {\em The Science of Complex Alloy
Phases} (TMS, Warrendale, 2005).

\bibitem{Urban04}
K.\ Urban and M.\ Feuerbacher, J. Non-Cryst. Solids {\bf 334}, 143 (2004)

\bibitem{Janot}
C.\ Janot, {\em Quasicrystals: A Primer} (Oxford University Press, Oxford
1997).

\bibitem{Zeng04}
X.\ Zeng, G.\ Ungar, Y.\ Liu, V.\ Percec, A.\ E.\ Dulcey, and J.\, K.\ Hobbs,
Nature {\bf 428}, 157 (2004).

\bibitem{Denton98}
A.\ R.\ Denton and H.\ L\"owen, Phys. Rev. Lett. {\bf 81}, 469 (1998).

\bibitem{LanconBillard}
F. Lan\c{c}on and L. Billard, Europhys. Lett. {\bf 2}, 625 (1986);
J. Phys. (France) {\bf 49}, 249 (1988).

\bibitem{Dzugutov92}
M.\ Dzugutov, Phys. Rev. A {\bf 46}, R2984 (1992); Phys. Rev. Lett. {\bf 70}, 2924 (1993)

\bibitem{Roth00}
J.\ Roth and A.\ R.\ Denton, Phys. Rev. E {\bf 61}, 6845 (2000).

\bibitem{Quandt99}
A.\ Quandt and M.\ P.\ Teter, Phys. Rev. B {\bf 59}, 8586 (1999).

\bibitem{Jagla98}
E.\ A.\ Jagla, Phys. Rev. E {\bf 58}, 1478 (1998).

\bibitem{Rechtsman06}
M.\ Rechtsman, F.\ Stillinger, and S.\ Torquato, Phys. Rev. E {\bf 73}, 011406
(2006).

\bibitem{RechtsmanDiverse}
M.\ C.\ Rechtsman, F.\ H.\ Stillinger, and S.\ Torquato, Phys. Rev. Lett. {\bf
95}, 228301 (2005); Phys. Rev. E {\bf 74}, 021404 (2006); Phys. Rev. E {\bf
75}, 031403 (2007).

\bibitem{Friedel}
J.\ Hafner, {\em From Hamiltonians to Phase Diagrams} (Sprin\-ger-Verlag, Berlin,
1987); J.\ A.\ Moriarty and M. Widom, Phys. Rev. B {\bf 56}, 7905 (1997).

\bibitem{Widom89}
M.\ Widom, D.\ P.\ Deng, and C.\ L.\ Henley, Phys. Rev. Lett. {\bf 63}, 310
(1989); K.\ J.\ Strandburg, L.-H.\ Tang, M.\ V.\ Jari{\'c},
Phys. Rev. Lett. {\bf 63}, 314 (1989).

\bibitem{Lee01}
A quasicrystal-crystal phase transition in a similar model has been predicted
by H.\ K.\ Lee, R.\ H.\ Swendsen, and M.\ Widom, Phys. Rev. B {\bf 64}, 224201
(2001).

\bibitem{Oxborrow93}
The elementary process is a zipper, see M.\ Oxborrow and C.\ L.\
Henley, Phys. Rev. B {\bf 48}, 6966 (1993).

\end{thebibliography}
\end{document}